\documentstyle[11pt,amssymb,epsf]{article}

\def\crampest{\medmuskip = 1mu plus 1mu minus 1mu}
\def\uncramp{\medmuskip = 4mu plus 2mu minus 4mu}
\def\ben{\begin{equation}}
\def\een{\end{equation}}

\let\a=\alpha    
    
 \let\m=\mu \let\n=\nu

\let\C=\Chi

\def\nn{\nonumber} \def\bd{\begin{document}} \def\ed{\end{document}}
\def\ds{\documentstyle} \let\fr=\frac \let\bl=\bigl \let\br=\bigr
\let\Br=\Bigr \let\Bl=\Bigl
\let\bm=\bibitem
\let\na=\nabla
\let\pa=\partial \let\ov=\overline
\newcommand{\be}{\begin{equation}}
\newcommand{\ee}{\end{equation}}
\def\ba{\begin{array}}
\def\ea{\end{array}}
\def\ft#1#2{{\textstyle{{\scriptstyle #1}\over {\scriptstyle #2}}}}
\def\fft#1#2{{#1 \over #2}}
\def\del{\partial}
\def\vp{\varphi}
\def\sst#1{{\scriptscriptstyle #1}}
\def\oneone{\rlap 1\mkern4mu{\rm l}}
\def\td{\tilde}
\def\wtd{\widetilde}
\def\ie{\rm i.e.\ }
\def\dalemb#1#2{{\vbox{\hrule height .#2pt
        \hbox{\vrule width.#2pt height#1pt \kern#1pt
                \vrule width.#2pt}
        \hrule height.#2pt}}}
\def\square{\mathord{\dalemb{6.8}{7}\hbox{\hskip1pt}}}
\newcommand{\ho}[1]{$\, ^{#1}$}
\newcommand{\hoch}[1]{$\, ^{#1}$}
\newcommand{\bea}{\begin{eqnarray}}
\newcommand{\eea}{\end{eqnarray}}
\newcommand{\ra}{\rightarrow}
\newcommand{\lra}{\longrightarrow}
\newcommand{\Lra}{\Leftrightarrow}
\newcommand{\ap}{\alpha^\prime}
\newcommand{\bp}{\tilde \beta^\prime}
\newcommand{\tr}{{\rm tr} }
\newcommand{\Tr}{{\rm Tr} }
\def\0{{\sst{(0)}}}
\def\1{{\sst{(1)}}}
\def\2{{\sst{(2)}}}
\def\3{{\sst{(3)}}}
\def\4{{\sst{(4)}}}
\def\5{{\sst{(5)}}}
\def\6{{\sst{(6)}}}
\def\7{{\sst{(7)}}}
\def\8{{\sst{(8)}}}
\def\n{{\sst{(n)}}}
\def\cA{{{\cal A}}}
\def\cF{{{\cal F}}}
\def\tV{\widetilde V}
\def\tW{\widetilde W}
\def\tH{\widetilde H}
\def\tE{\widetilde E}
\def\tF{\widetilde F}
\def\tA{\widetilde A}
\def\im{{{\rm i}}}
\def\tY{{{\wtd Y}}}
\def\ep{{\epsilon}}
\def\vep{{\varepsilon}}
\def\R{\rlap{\rm I}\mkern3mu{\rm R}}
\def\bD{{{\bar D}}}

\def\R{\rlap{\rm I}\mkern3mu{\rm R}}
\def\bD{{{\bar D}}}
\def\R{{{\Bbb R}}}
\def\C{{{\Bbb C}}}
\def\H{{{\Bbb H}}}
\def\CP{{{\Bbb C}{\Bbb P}}}
\def\RP{{{\Bbb R}{\Bbb P}}}
\def\Z{{{\Bbb Z}}}
\def\bA{{{\Bbb A}}}
\def\bB{{{\Bbb B}}}
\def\bC{{{\Bbb C}}}
\def\bD{{{\Bbb D}}}
\def\bE{{{\Bbb E}}}
\def\bZ{{{\Bbb Z}}}
\def\Re{{{\frak{Re}}}}
\def\Im{{{\frak{Im}}}}
\def\cosec{{\,\hbox{cosec}\,}}
\def\Gm{{\Gamma_{\!\! -}}}
\def\Gp{{\Gamma_{\!\! +}}}
\def\stan{{standard }}
\def\nonstan{{supernumerary }}
\def\btheta{{\bar\theta}}

\newcommand{\tamphys}{\it Center for Theoretical Physics,
Texas A\&M University, College Station, TX 77843, USA}
\newcommand{\umich}{\it Michigan Center for Theoretical Physics,
University of Michigan\\ Ann Arbor, MI 48109, USA}
\newcommand{\upenn}{\it Department of Physics and Astronomy,
University of Pennsylvania\\ Philadelphia,  PA 19104, USA}
\newcommand{\SISSA}{\it  SISSA-ISAS and INFN, Sezione di Trieste\\
Via Beirut 2-4, I-34013, Trieste, Italy}

\newcommand{\ihp}{\it Institut Henri Poincar\'e\\
  11 rue Pierre et Marie Curie, F 75231 Paris Cedex 05}

\newcommand{\damtp}{\it DAMTP, Centre for Mathematical Sciences,
 Cambridge University\\ Wilberforce Road, Cambridge CB3 OWA, UK}
\newcommand{\itp}{\it Institute for Theoretical Physics, University of
California\\ Santa Barbara, CA 93106, USA}

\newcommand{\auth}{M. Cveti\v{c}\hoch{\dagger},
H. L\"u\hoch{\star} and C.N. Pope\hoch{\ddagger}}

\begin{document}
\begin{flushright}
\hfill{CTP TAMU-05/02}\ \ \ {UPR-985-T}\ \ \
{MCTP-02-18}\\
{March 2002}\ \ \
{hep-th/0203229}
\end{flushright}


\begin{center}
{ \large {\Large\bf M-theory PP-Waves, Penrose Limits and \\  
                    Supernumerary Supersymmetries}}

\vspace{20pt}
\auth

\vspace{3pt}
{\hoch{\dagger}\upenn}

\vspace{3pt}


\vspace{3pt}
{\hoch{\star}\umich}

\vspace{3pt}
{\hoch{\ddagger}\tamphys}

\vspace{3pt}

\underline{ABSTRACT}
\end{center}

   We study supersymmetric pp-waves in M-theory, their dimensional
reduction to D0-branes or pp-waves in type IIA, and their
T-dualisation to solutions in the type IIB theory.  The general class
of pp-waves that we consider encompass the Penrose limits of
AdS$_p\times S^q$ with $(p,q)=(4,7),\, (7,4),\, (3,3),\, (3,2),\,
(2,3),\, (2,2)$, but includes also many other examples that can again
lead to exactly-solvable massive strings, but which do not arise from
Penrose limits.  All the pp-waves in $D=11$ have 16 ``standard''
Killing spinors, but in certain cases one finds additional, or
``supernumerary,'' Killing spinors too.  These give rise to
linearly-realised supersymmetries in the string or matrix models.  A
focus of our investigation is on the circumstances when the Killing
spinors are independent of particular coordinates ($x^+$ or
transverse-space coordinates), since these will survive at the
field-theory level in dimensional reduction or T-dualisation.

{\vfill\leftline{}\vfill
\vskip 5pt
\footnoterule
{\footnotesize \hoch{\dagger} Research supported in part by DOE grant
DE-FG02-95ER40893 and NATO grant 976951. \vskip -12pt} \vskip 14pt
{\footnotesize \hoch{\star} Research supported in full by DOE grant
DE-FG02-95ER40899 \vskip -12pt} \vskip 14pt
{\footnotesize  \hoch{\ddagger} Research supported in part by DOE
grant DE-FG03-95ER40917.\vskip  -12pt}}

\pagebreak
\setcounter{page}{1}


\section{Introduction}\label{introsec}

    The Penrose limit \cite{penrose} of the AdS$_5\times S^5$ solution
of type IIB theory is a pp-wave with maximal supersymmetry
\cite{blafighulpap,blafighulpap2}.  This result is of considerable
interest within the framework of the AdS/CFT correspondence, since the
pp-wave provides a background for which the string theory action in
light-cone gauge describes a massive free string, which is exactly
solvable \cite{met,bermalnas}, thus allowing explicit comparisons with
results in the dual gauge theory \cite{bermalnas}.  The Penrose limit
of the AdS$_4\times S^7$ and AdS$_7\times S^4$ solutions of M-theory
also gives rise to a maximally-supersymmetric pp-wave, obtained in
\cite{kg}, and this provides a simple background for the DLCQ
description of M-theory, and the corresponding matrix-model in this
regime \cite{bermalnas}.  Subsequent papers have explored a variety of
consequences and generalisations of these observations
\cite{bur1}-\cite{bur10}.

     In a previous paper \cite{clppenrose}, we studied a wider class
of supersymmetric pp-waves in the type IIB theory, generalising the
maximally-supersymmetric one that arises as the Penrose limit of
AdS$_5\times S^5$.  In particular, we allowed for a more general
structure of the constant self-dual five-form field strength; these
structures were motivated by the flat (orbifold) limit of a special
holonomy transverse space for the pp waves.  In fact any pp-wave
within the general class automatically has 16 Killing spinors, which
we therefore denoted as ``standard'' Killing spinors.  In special
cases one finds that there can be additional Killing spinors, which we
denoted as ``supernumerary'' Killing spinors. The maximum number, 16,
of these is achieved for the Penrose limit of AdS$_5\times S^5$.  (In
this class we also found another example of the Penrose limit of
AdS$_3\times S^3$ arising form an D3/D3-intersection.)

   The focus of our study in \cite{clppenrose} was to determine the
circumstances under which one obtains supernumerary Killing spinors in
the type IIB pp-waves.  These are important when one considers the
exactly-solvable string models in the pp-wave backgrounds; we found
that it is the supernumerary Killing spinors that are in one to one
correspondence with the associated linearly-realised worldsheet
supersymmetries of the corresponding string action.  In fact the
string theory is solved by going to the light-cone gauge, with the
$x^+$ coordinate in the pp-wave being set equal to the world-sheet
time coordinate.  In order that the linearly-realised world-sheet
supersymmetries be unbroken, it is necessary therefore that the
associated supernumerary Killing spinors be independent of the
coordinate $x^+$, which is indeed the case in all the type IIB pp-waves.  For
instance, all 16 supernumerary Killing spinors in the Penrose limit of
AdS$_5\times S^5$ have this property \cite{bermalnas,clppenrose}.

   A further significance of having Killing spinors in the type IIB
pp-waves that are independent of $x^+$ is that after performing a
T-duality transformation on the $x^+$ coordinate (which is always a
Killing direction), the resulting type IIA solution will also be
supersymmetric.  It can be lifted to M-theory, where it acquires an
interpretation as a supersymmetric deformed M2-brane, \ie an M2-brane
in which an additional 4-form flux is turned on in the transverse
space \cite{hawtay}-\cite{newspin7}.  An intriguing feature of the
deformed M2-branes obtained by this T-dualisation procedure is that if
any of the Killing spinors originate from {\it supernumerary} Killing
spinors (which are $x^+$-independent), then in the M-theory picture
they solve the Killing-spinor equations despite violating the
criterion that is usually applied \cite{2bec,hawtay,kbec} for testing
whether a supersymmetry survives when the extra 4-form flux is turned
on \cite{clppenrose}.

   In this paper, we study supersymmetric pp-waves in M-theory.  In
particular, we allow for rather general structures for the constant
4-form field strength of M-theory, motivated from the flat (orbifold)
limit of special holonomy transverse space for the pp-waves.  These
possible structures fall into two classes. Focusing on the nature of
supersymmetry, we again find that there are always 16 ``standard''
Killing spinors, and that additional ``supernumerary'' Killing spinors
can arise in special cases.  Unlike the case of pp-waves in type IIB,
however, it is no longer automatic that supernumerary Killing spinors
are independent of $x^+$.  The Penrose limit of AdS$_4\times S^7$ or
AdS$_7\times S^4$ provides the unique example where the number of
supernumerary Killing spinors attains its maximum, namely 16
\cite{kg}.  Unlike the maximally-supersymmetric pp-wave in type IIB,
however, here in the M-theory maximally-supersymmetric pp-wave the 16
supernumerary Killing spinors all depend on $x^+$.

   Again our focus is on the occurrence of supernumerary Killing
spinors, and also on determining the dependence of all the Killing
spinors on the $x^+$ coordinate and the 9 transverse coordinates
$z^i$.  These dependences are of importance when one considers
reductions to type IIA, and subsequent T-dualisation to type IIB,
since they determine whether there will be supersymmetries 
in the type IIA and IIB supergravity solutions.

    Depending upon whether one reduces from $D=11$ on $x^+$ or on one
of the transverse coordinates $z^i$, one either obtains a D0-brane or
a pp-wave in the type IIA theory.  In the case of a pp-wave, the
string theory in this background is again exactly-solvable by going to
the light-cone gauge, giving rise to a free massive theory.  In the
case of a reduction instead on $x^+$, the D0-brane world-particle
action leads to a DLCQ description of the M-theory matrix model in
this sector.  Thus these backgrounds of M-theory provide for dual
descriptions, either in terms of a solvable type IIA string action or
in a matrix theory model. In particular, the supernumerary
supersymmetry plays a key role in determining the supersymmetry of the
string action as well as supersymmetry of the matrix model.

   We begin in section 2 by setting up our formalism for the pp-waves
in M-theory, and obtaining the criterion for the existence of Killing
spinors.  We then study the coordinate dependences of the Killing
spinors, for the various choices of constant 4-form fluxes in the flat
nine-dimensional transverse space that we are considering.  This allows
us to discuss the supersymmetry of the type IIA D0-branes and pp-waves
that we can obtain by dimensional reduction.  We also show how some of
our pp-wave solutions are related to Penrose limits of M2-brane and
M5-brane intersections.

   In section 3 we derive the light-cone action for type IIA strings
in arbitrary bosonic backgrounds, making use of earlier covariant
results for the Green-Schwarz action up to quadratic order in fermions
that were obtained in \cite{clps}.  We also obtain an analogous result
for the light-cone action for the type IIB string, again valid for
arbitrary bosonic backgrounds, and based on covariant results obtained
in \cite{clps}.  Using these light-cone actions, we study the
properties of some of the pp-wave solutions that we obtain by
dimensional reduction and T-duality.

   In certain cases the M-theory pp-waves have no isometry direction
within the nine-dimensional transverse space, and so they cannot be
reduced to give pp-waves in ten dimensions.  Under these
circumstances, where the pp-waves are intrinsically
eleven-dimensional, one is led to considering a DLCQ description in
this background, leading to a matrix model.  In particular, we derive
the matrix-model action, whose supersymmetry is in one to one
correspondence with the supernumerary supersymmetries of the
corresponding pp-wave background.

\section{Supersymmetry of pp-waves in M-theory}

\subsection{General formalism}

    We shall consider pp-wave solutions of $D=11$ supergravity, where
the metric and 4-form are given by
\bea
ds_{11}^2 &=& -4 dx^+\, dx^- + H\, d{x^+}^2 + dz_i^2\,,\label{hmet}\\
F_\4 &=& \mu\, dx^+\wedge \Phi_\3\,,\label{f4exp}
\eea
where $\Phi_\3$ is a harmonic 3-form in the flat nine-dimensional
transverse space whose metric is $dz_i^2$, $\mu$ is a constant, and we
are taking $H$ here to depend only on $z^i$.  In the vielbein basis
$e^+=dx^+$, $e^- = -2dx^- + \ft12 H\, dx^+$, $e^i=dz^i$, for which the
metric is $ds_{11}^2= 2 e^+\, e^- + e^i\, e^i$, the spin connection is
given by
\be
\omega_{+i}= \ft12 \del_i H\, e^+\,,\qquad \omega_{-i}=\omega_{+-}=
\omega_{ij}=0\,,\label{spincon}
\ee
and the only non-vanishing Riemann tensor components, in the vielbein basis,
are
\be
R_{+i+j}=-\ft12 \del_i\del_j H\,.
\ee
This implies that the only non-vanishing Ricci-tensor component is
$R_{++}=-\ft12 \square H$.  The $D=11$ supergravity equations are
therefore satisfied if $H$ obeys the equation
\be
\square\, H = -\ft16\, \mu^2\, |\Phi_\3|^2\,.\label{heq}
\ee
In this paper we shall focus on the cases where $\Phi_\3$ is a 
covariantly-constant 3-form.  It is sufficient for our purposes to take the
solution for $H$ to be 
\be
H = c_0 + \fft{Q}{r^7} - \sum_i \mu_i^2\, z_i^2\,,\label{hsol}
\ee
where $c_0$, $Q$ and $\mu_i$ are constants, and $r^2\equiv z_i\, z_i$.  
It follows from (\ref{heq}) that the $\mu_i$ are subject to the condition
\be
\sum_i \mu_i^2 =\ft1{12}\, \mu^2\, |\Phi_\3|^2\,.\label{muicon}
\ee
When $\mu=0$ (and hence $\mu_i=0$), the solution becomes a standard 
pp-wave in $D=11$, whose dimensional reduction gives a D0-brane in the
type IIA theory, and the pp-wave charge $Q$ becomes the charge of the
D0-brane.

   The supercovariant derivative appearing in the
supersymmetry transformation rule $\delta\psi_M = D_M\, \ep$ is given by  
\be
D_M = \nabla_M - \ft1{288}\,(\Gamma_M{}^{N_1\cdots N_4}\, 
      F_{N_1\cdots N_4} - 8 F_{M N_1\cdots N_3}\, \Gamma^{N_1 \cdots N_3})
\,.
\ee
Defining $D_M=\nabla_M + \Omega_M$, we therefore find
\bea
&&\nabla_+ = \del_+ +\ft14 \del_i H\, \Gm\, \Gamma_i\,,\qquad
\nabla_-=\del_-\,,\qquad \nabla_i = \del_i\,,\nn\\
&&\Omega_+= -\ft{\im}{12}\mu\, (1+\Gamma_-\, \Gamma_+)\, W\,,
\qquad \Omega_- = 0\,,\qquad
\Omega_i = \ft{\im}{24}\, \mu\, \Gm\, (\Gamma_i\, W + 3W\, \Gamma_i)\,,
\label{supercov}
\eea
where we have defined
\be
W \equiv \ft{\im}6 \Phi_{ijk}\, \Gamma_{ijk}\,.\label{wdef}
\ee

    It follows immediately from (\ref{supercov}) that Killing spinors $\ep$,
satisfying $D_M\, \ep=0$, are independent of $x^-$.  Since 
$\Omega_i\, \Omega_j=0$ we have $\del_i\, \del_j\, \ep=0$ and hence it follows
that 
\be
\ep = (1 - z^i\, \Omega_i)\, \chi\,,\label{zidep}
\ee
where $\chi$ depends only on $x^+$.  Finally, from $D_+\, \ep=0$ one deduces
that
\be
\del_+\, \chi - \ft{\im}{12}\mu\, (\Gm\, \Gp + 1)\, W\, \chi=0\label{xpdep}
\ee
and
\be
\Big[\mu^2\,z^i\,  (\Gamma_i\, W^2 + 9 W^2\, \Gamma_i + 6 W\, \Gamma_i\, W)
 + 72 \del_i H\, \Gamma_i \Big]\, \Gm\, \chi=0\,.\label{susyeq}
\ee
Thus (\ref{susyeq}) determines the number of Killing spinors, while
(\ref{zidep}) and (\ref{xpdep}) determine their $z^i$ and $x^+$ dependence.

    Since $\Gp\, \Gm + \Gm\, \Gp=2$ and $\Gp^2=\Gm^2=0$, we have
a unique decomposition $\chi= \chi_+ + \chi_-$ for any $\chi$, where
$\chi_+\equiv \ft12 \Gp\, \Gm\, \chi$ and $\chi_-\equiv \ft12 
\Gm\, \Gp\, \chi$ have the defining properties $\Gp\, \chi_+=0$ and
$\Gm\, \chi_-=0$.

   It is evident from (\ref{susyeq}) that there will always be 16
Killing spinors corresponding to $\chi=\chi_-$.  Accordingly, we refer
to these as ``standard Killing spinors,'' since they exist for any
choice of the function $H$ that satisfies the field equation
(\ref{heq}).  In particular, the pp-wave charge $Q$ can be non-zero.
In certain cases there can also be additional Killing spinors
corresponding to $\chi=\chi_+$.  We refer to these as ``supernumerary
Killing spinors.''  We shall construct the two categories of Killing
spinors, and discuss their coordinate dependences, in the following
two subsections.

    Before discussing the two categories of Killing spinors, let us be a
little more specific in our choice of constant 3-forms $\Phi_\3$.  It turns 
out to be natural to restrict attention to cases where the associated
matrix $W$, defined by (\ref{wdef}), is a sum of individual terms $W_\a$
that all commute with each other.  This can be seen to lead to two
inequivalent maximal sets of terms, which we shall refer to as Case 1 and 
Case 2.  For the two cases we have
\bea
\noindent{\underline{{\bf Case 1}}}:&&
\phantom{why doesn't this do noindent?????????????????????????????????
 ?????????????}
\nn\\
&&\Phi_\3 = m_1\, dz^{129} + m_2\, dz^{349} + m_3\, dz^{569} +
           m_4\, dz^{789}\,,\label{case1}\\
&&\nn \\
\noindent{\underline{{\bf Case 2}}}:&&
\phantom{why doesn't this do noindent?????????????????????????????????
 ?????????????}
\nn\\
&&\Phi_\3= m_1\, dz^{123} + m_2\, dz^{145} + m_3\, dz^{167} +
           m_4\, dz^{246}\nn\\
&& \qquad \qquad + 
m_5\, dz^{257} + m_6\, dz^{347} +m_7\, dz^{356} \,,\label{case2}
\eea
where we have defined $dz^{ijk}\equiv dz^i\wedge dz^j\wedge dz^k$.  It
should be noted that unless all four of the $m_\a$ coefficients
are non-zero in Case 1, it is in fact encompassed (after a relabelling
of coordinates) within Case 2.

    It is straightforward to verify that if we construct $W$ as in
(\ref{wdef}), and write it as
\be
W = \sum_\a m_\a\, W_\a\,,\label{wsum}
\ee
where $W_\a$ denotes the individual $\Gamma_{ijk}$ structures (for
example $W_1=\im\, \Gamma_{129}$ is one of the four structures in Case
1), then we shall have $[W_\a,W_\beta]=0$.  In consequence, for either
Case 1 or Case 2 we can choose a basis for the gamma matrices in which
the $W_\a$ are all diagonal.  It is useful to have in mind such a
diagonal choice of basis in the subsequent discussion.
 
  For our canonical choices, one can see that if the $m_\a$ are taken equal
then $\Phi_\3$ in Case 1 can be expressed as $m\, dz_9\wedge J$, where
$J$ is the K\"ahler form for the eight-dimensional flat space with 
metric $\sum_{i=1}^8 dz_i^2$.  Likewise, if the $m_\a$ are set equal in
Case 2, $\Phi_\3$ can be expressed as $m\, \Psi_\3$, where $\Psi_\3$ is
a $G_2$-invariant associative 3-form in the flat seven-dimensional space
with metric $\sum_{i=1}^7 dz_i^2$.   

\subsection{The 16 standard Killing spinors}\label{stansec}

   The 16 standard Killing spinors correspond to taking
$\chi=\chi_-$, \ie they are defined by $\Gm\, \chi=0$.  It is evident
from (\ref{supercov}) and (\ref{zidep}) that they are all independent
of all of the $z^i$ coordinates.  It is also evident from
(\ref{xpdep}) that they will have $x^+$ dependence given by
\be
\chi =e^{\fft{\im}4\, \mu\, x^+\, W}\, \chi_0\,,
\ee
where $\chi_0$ is any constant spinor satisfying $\Gm\, \chi_0=0$.  If 
$W$ annihilates any of these spinors, then the associated 
``standard'' Killing spinor will be independent of $x^+$ (and so, in
fact, it will be independent of all the coordinates).  The discussion now
divides into two, according to whether we take $\Phi_\3$ to be given by
(\ref{case1}) or (\ref{case2}):

\bigskip
\noindent{\underline{\bf Case 1}}:
\medskip

   For $\Phi_\3$ given by (\ref{case1}), the eigenvalues of $W$ are
\be
\lambda_i = \pm m_1 \pm m_2 \pm m_3 \pm m_4\,,\label{case1wev}
\ee
where the $\pm$ choices are all independent. Each eigenvalue occurs twice, 
making the 32 in total.  In the subspace of eigenspinors annihilated by 
$\Gm$ one gets each eigenvalue once, and likewise in the subspace 
annihilated by $\Gp$.   For the standard Killing spinors arising in 
Case 1, it therefore follows that the possible numbers that are independent of
$x^+$ can be $N_{\rm stan}=$ 0, 2, 4 or 8.  

   The number $N_{\rm stan}=0$ is achieved for generic choices of the
$m_\a$; $N_{\rm stan} =2$ is achieved for choices where
$m_1+m_2+m_3+m_4=0$; $N_{\rm stan}=4$ is achieved for choices with $m_4=0$
and $m_1+m_2+m_3=0$ (or permutations); and $N_{\rm stan}=8$ is achieved
for choices with $m_3=m_4=0$ and $m_1+m_2=0$ (or permutations).

\vfill\eject
\bigskip
\noindent{\underline{\bf Case 2}}:
\medskip

   When $\Phi_\3$ is given by (\ref{case2}), we find that again $W$
generically has sixteen different eigenvalues, each occurring twice.
One copy of the sixteen again occurs in each of the $\Gm$ and $\Gp$
subspaces.  The eigenvalues are given by $\pm\lambda_i$, where
\be
\lambda_8= m_1 + m_2 + m_3 - m_4 + m_5 + m_6 + m_7\,,\label{case2wev}
\ee
and $\lambda_i$ for $1\le i\le 7$ is given by reversing the sign of
each $m_\a$ that occurs as a coefficient of any term containing the
gamma matrix $\Gamma_i$.  The numbers of standard Killing spinors that
are independent of $x^+$ that can be achieved for these Case 2
examples are therefore $N_{\rm stan}=2n$, where $0\le n\le 6$ is the
number of $\lambda_i=0$ that are arranged to vanish by choosing the
$m_\a$ appropriately.  Thus for Case 2 we can have $N_{\rm stan}=$ 0,
2, 4, 6, 8, 10 or 12 standard Killing spinors that are independent of
$x^+$.

\subsection{Supernumerary Killing spinors}\label{supernumsec}

    We now turn to the discussion of supernumerary Killing spinors,
for which $\Gp\, \chi=0$.  In a generic pp-wave solution there will be
none of these, but they can arise in special cases when $H$ given in
(\ref{hsol}) is quadratic in $z^i$ (\ie the pp-wave charge $Q=0$), and
the distribution of $\mu_i$ coefficients (which must in any case
satisfy (\ref{muicon})) is chosen appropriately.  The numbers of
supernumerary Killing spinors that can be achieved depends also on the
choice of $\Phi_\3$.

   The equation (\ref{susyeq}) that determines the number of Killing spinors
admits solutions for supernumerary Killing spinors ($\Gp\, \chi=0$) if
$H$ in (\ref{hsol}) depends on $z^i$ only quadratically, and
\be
\Big[\mu^2\, (\Gamma_i\, W^2 + 9 W^2\, \Gamma_i + 6 W\, \Gamma_i\, W) 
 - 144 \mu_i^2\, \Gamma_i\Big]\, \chi=0\,,\label{susyeq2}
\ee
where $\Gp\, \chi=0$.  

   As we discussed when we made our choices (\ref{case1}) or
(\ref{case2}) for $\Phi_\3$, the resulting terms $W_\a$ defined by
(\ref{wdef}) and (\ref{wsum}) can all be simultaneously diagonalised
(for each of Case 1 or Case 2 separately), by means of an appropriate
similarity transformation of the gamma matrices.  It is convenient to
assume that such a basis for the gamma matrices has been chosen.

   With respect to a diagonal basis, it is evident that
\be
X_i\equiv \Gamma_i\, W\, \Gamma_i + 3W\label{xdef}
\ee
is also diagonal, and therefore that (\ref{susyeq2}) can be rewritten as
\be
(\mu^2\, X_i^2 - 144 \mu_i^2)\, \Gamma_i\, \chi=0\,,
\qquad \hbox{for each $i$}\,.\label{snnum}
\ee

   From (\ref{zidep}), it now follows that the solutions of (\ref{snnum}) 
will give the supernumerary Killing spinors, with $z^i$ dependence given
by
\be
\ep = (1-\ft{\im}{2}\, \mu_i\, z^i\, \Gm\, \Gamma_i)\, \chi\,.
\label{zidep2}
\ee
In particular, this means that a supernumerary Killing spinor is
independent of a given coordinate $z^i$ if and only if the associated
coefficient $\mu_i$ in (\ref{hsol}) is zero.

   The discussion of the supernumerary Killing spinors now divides
into the two possibilities for $\Phi_\3$, given by (\ref{case1}) or
(\ref{case2}).

\bigskip
\noindent{\underline{\bf Case 1}}:
\medskip

   In this case $\Phi_\3$ is given by (\ref{case1}).  In the direction
$i=9$ we have $X_9=4W$, and so $\mu_9^2 = \ft1{9}\, \mu^2\,
\lambda^2$, where $\lambda$ is one of the eigenvalues of $W$ given in
(\ref{case1wev}).  Without loss of generality, since the other
eigenvalues differ only in sign permutations of the $m_\a$, we can
take
\be
\mu_9^2 = \ft19 \mu^2\, (m_1+m_2+m_3+m_4)^2\,.\label{case1mu9}
\ee
The remaining $\mu_i$ for $1\le i\le 8$ are then given by
\bea
\mu_1^2=\mu_2^2 &=& \ft1{36}\, \mu^2\, (-2m_1+m_2+m_3+m_4)^2\,,\nn\\
\mu_3^2=\mu_4^2 &=& \ft1{36}\, \mu^2\, (m_1-2m_2+m_3+m_4)^2\,,\nn\\
\mu_5^2=\mu_6^2 &=& \ft1{36}\, \mu^2\, (m_1+m_2- 2m_3+m_4)^2\,,\nn\\
\mu_7^2=\mu_8^2 &=& \ft1{36}\, \mu^2\, (m_1+m_2+m_3-2 m_4)^2\,.
\label{muisol}
\eea
To see this, we note that if $(X_9-\kappa_9)\, \Gamma_9\, \chi=0$ then
$4W\, \chi = \kappa_9\, \chi$.  We are taking
$\kappa_9=4(m_1+m_2+m_3+m_4)$.  Substituting into $(X_1-\kappa_1)\,
\Gamma_1\, \chi =0$ we therefore find $\ft14 \kappa_9\, \chi + 3 \wtd
W\, \chi -\kappa_1\, \chi=0$, where $\wtd W= \Gamma_1\, W\, \Gamma_1$.
From (\ref{case1}) and (\ref{wdef}) it follows that the diagonal
matrix $\wtd W$ has eigenvalues that are just those of $W$ but with
$m_2$, $m_3$ and $m_4$ reversed in sign, and so $\wtd W\, \chi=
(m_1-m_2-m_3-m_4)\, \chi$.  Thus we deduce that $\kappa_1=
2(2m_1-m_2-m_3-m_4)$.  Applying an analogous argument for each
direction $i$, we arrive at (\ref{muisol}).

   For a generic choice of the constants $m_\a$, there are precisely
two supernumerary Killing spinors.  This is because a given bosonic
solution has fixed values for the coefficients $\mu_\a$, and so there
are two solutions to (\ref{susyeq2}) since there is a twofold
degeneracy in (\ref{muisol}).  In special cases, where the $m_\a$ are
chosen so that two or more of the expressions in (\ref{muisol}) are
equal, there can therefore be more solutions of (\ref{susyeq2}).  It
is an elementary exercise to enumerate all the possible numbers of
supernumerary supersymmetries that can be achieved for specific
choices of $m_\a$.  As in the case of type IIB pp-wave solutions, the
supernumerary supersymmetries can lead to a variety of
``non-standard'' fractions of total supersymmetry that exceed $\ft12$
\cite{clppenrose}.

    It is worth remarking that for Case 1, as a consequence of the
equation $X_9=4W$, it follows that supernumerary Killing spinors are
independent of $x^+$ if and only if they are independent of $z_9$.

\bigskip
\noindent{\underline{\bf Case 2}}:
\medskip

   When the 3-form $\Phi_\3$ is given by (\ref{case2}), it follows
that we shall have $X_8=X_9=2W$, and so from (\ref{snnum}) we
shall have $\mu_8^2=\mu_9^2=\ft1{36}\, \mu^2\, \lambda^2$, where
$\lambda$ is one of the eigenvalues of $W$.  These are 
now given by $\lambda_8$ in (\ref{case2wev}), together with
$\lambda_i$ for $1\le i\le 7$ as described below (\ref{case2wev}).
Without loss of generality, since the $\m_\a$ have not yet been 
specified, we may choose
\be
\mu_8^2=\mu_9^2 = \ft1{36}\mu^2\, \lambda_8^2 = \ft1{36} \mu^2\, 
(m_1+m_2+m_3-m_4 + m_5 + m_6+ m_7)^2\,.
\ee
For the other constants $\mu_i$, we shall therefore have
\be
\mu_i^2 = \ft1{144}\, \mu^2\, (\lambda_8- 3\lambda_i)^2\,,\qquad
1\le i\le 7\,.
\ee
where $\lambda_8$ and $\lambda_i$ are given by (\ref{case2wev}) and
below.  For example, we shall have $\mu_1^2= \ft1{36} \mu^2\, (2 m_1 +
2 m_2 + 2 m_3 + m_4 - m_5 - m_6 - m_7)^2$.

\subsection{Supersymmetry of type IIA pp-waves and D0-branes}

      Having obtained the M-theory pp-waves, we can dimensionally
reduce the solutions to $D=10$, giving rise to D0-branes if we reduce
on the $x^+$ coordinate, or to type IIA pp-waves if we reduce instead
on any of the $z^i$ coordinates.  Of course a reduction on a
particular $z^i$ coordinate is possible only if it is a Killing
direction, which means that the associated coefficient $\mu_i$ in the
metric function $H$ must vanish.\footnote{There are, of course, many
other Killing vectors in the pp-wave metric, which could be used for
Kaluza-Klein reduction, but we are not considering these here (see,
however, section \ref{explicitsec}).
Some examples are discussed in \cite{mich}; these typically give
rise to an extra (constant) flux in the lower dimension, coming from
a non-vanishing Kaluza-Klein vector potential.}

       First let us consider a reduction on $x^+$.  This is a Killing
direction for all pp-wave solutions.  However, some or all of the
Killing spinors in a given solution may be dependent on the $x^+$
coordinate, in which case they will not survive in the reduction of
the solution to type IIA supergravity.  As we saw from (\ref{xpdep}),
the criterion for a Killing spinor to be independent of $x^+$ is that
it should be annihilated by $W$.  For the standard Killing spinors,
the fraction of the 16 Killing spinors that will survive the reduction
on $x^+$ depends on the detailed structure of $W$.  The 16 standard
Killing spinors exist for any solution for $H$, subject to
(\ref{heq}).  In particular, they exist when the D0-brane charge $Q$
is turned on.  The supernumerary Killing spinors, on the other hand,
are all eigenvectors of $W$ with the same eigenvalue, and hence they
are all $x^+$-independent if $W\chi=0$, but $x^+$-dependent if $W\,
\chi\ne0$.  The supernumerary Killing spinors exist only if $Q=0$ 
and the $\mu_i$ are distributed appropriately.

     For a reduction on one of the transverse coordinates $z^i$, the
the corresponding constant $\mu_i$ in the expression for $H$ in the
metric (\ref{hmet}) must vanish, in order that $\del/\del z^i$ be a
Killing vector.  As we saw in (\ref{zidep2}), the Killing spinors are
then also independent of the coordinate $z^i$, and hence they will all
survive in the reduction.  In section \ref{stansec}, it was observed
that the 16 standard Killing spinors are all independent of $z^i$.

   When $\Phi_\3$ is contained within the Case 1 in (\ref{case1}), 
the direction $z_9$ is singled out.  It was observed in the discussion
of Case 1 in section \ref{supernumsec} that if $\mu_9=0$ then  
we have also $W\chi=0$, and so this implies that if $z_9$ is a Killing
direction then the supernumerary Killing spinors will not only
be independent of $z_9$, but also of $x^+$.

  In Case 2, where $\Phi_\3$ is given by (\ref{case2}), the directions
$z_8$ and $z_9$ are singled out.  If we arrange for $\mu_8=\mu_9=0$
(they are always equal), then as shown in \ref{supernumsec} we also
have $W\chi=0$, and so the supernumerary Killing spinors will be
independent of $x^+$ as well as the reduction coordinate $z_8$ or
$z_9$.

   In fact these various reductions to type IIA can be related to the
general type IIB pp-waves obtained in \cite{clppenrose}, by means of
T-duality.  If we dimensionally reduce on $z_9$ in Case 1, or on $z_8$
or $z_9$ in Case 2, which is possible if parameters are chosen so that
the corresponding $\mu_i$ coefficient vanishes, the resulting
supernumerary Killing spinors are all independent $x^+$.  This implies
that the type IIA string action then has linearly-realised
supersymmetries.

    We can also obtain large classes of type IIA pp-wave solutions in
which other $\mu_i$ parameters are instead zero.  (That is to say,
$\mu_i$ other than $\mu_9$ in Case 1, or $\mu_8=\mu_9$ in Case 2.)  In
these circumstances, there can exist supernumerary Killing spinors
that are dependent on $x^+$.  One would then obtain a type IIA
solution where some, or all, of the world-sheet supersymmetries were
non-linearly realised.

\subsection{An explicit example}\label{explicitsec}

    Here we consider an explicit example where only $m_1$ and $m_2$
are non-vanishing.  (This can equally well be for either Case 1 in
(\ref{case1}) or Case 2 in (\ref{case2}), since the two are then
equivalent after coordinate relabellings.)   Taking $m_1$ and $m_2$ as 
given, we can then arrange for two choices for the $\mu_i$ coefficients that
will give rise to supernumerary Killing spinors.  These are
summarised in the following table:

\bigskip\bigskip
\centerline{
\begin{tabular}{|c|c|c|c|c|}\hline
$\mu_1^2=\mu_2^2$ & $\mu_3^2=\mu_4^2$ & $\mu_5^2=\mu_6^2$ &
$\mu_7^2=\mu_8^2$ & $\mu_9^2$ \\ \hline\hline
$\ft1{36}(-2m_1+m_2)^2$ & $\ft1{36} (m_1 -2m_2)^2$ & 
$\ft1{36}(m_1+m_2)^2$ & $\ft{1}{36}(m_1+m_2)^2$ &
$\ft1{9}(m_1+m_2)^2$\\ \hline
$\ft1{36}(2m_1+m_2)^2$ & $\ft1{36} (m_1 +2m_2)^2$ & 
$\ft1{36}(m_1-m_2)^2$ & $\ft{1}{36}(m_1-m_2)^2$ &
$\ft{1}{9}(m_1-m_2)^2 $\\ \hline
\end{tabular}}
\bigskip

\centerline{Table 1: Choices for $\mu_i$ with fixed $m_1$ and $m_2$
 that give supernumerary Killing spinors}
\bigskip\bigskip

    The second choice is nothing but the first, with one of the two $m_i$
reversed in sign.  However, if the $m_\a$ are given, fixed parameters, 
then these two choices for the $\mu_\a$ correspond to two independent
solutions.\footnote{In our previous discussion, we adopted an ``active'' 
viewpoint when discussing the possible occurrences of supernumerary
Killing spinors, rather than the ``passive'' viewpoint we are adopting here.
Namely, we previously took a fixed choice for how the eigenvalues 
were to be expressed in terms of the $m_\a$, and then covered the spectrum of
possibilities by allowing the $m_\a$ to be chosen freely.  The two
viewpoints are clearly equivalent, if appropriate care is taken.}

      For generic but fixed values of $m_\a$, each choice in Table
1 gives rise to 8 supernumerary Killing spinors.  When $m_1=m_2$ in
the second choice, the 8 supernumerary Killing spinors are all
independent of $x^+$.  This then corresponds to the Penrose limit of
the M2/M5-brane system (AdS$_3\times S^3 \times T^4$).  By contrast,
in the first choice in Table 1 the 8 supernumerary Killing spinors
depend on the $x^+$ coordinate.  For both choices, 8 of the 16 standard Killing
spinors are independent of $x^+$.  Another special case arises when either
$m_1=\pm 2m_2$ or $m_2=\pm 2m_1$, which implies that $\mu_1=\mu_2=0$ or
$\mu_3=\mu_4=0$.  Interestingly enough, this case is
T-dual to the maximally supersymmetric pp-wave arising from the
Penrose limit of AdS$_5\times S^5$.  Suppose, for example, we have the 
choice giving $\mu_1=\mu_2=0$.  We can then reduce on $z_1$ and
T-dualise on $z_2$.  The resulting type IIB solution is the 
maximally-supersymmetric pp-wave, in a slightly non-standard
coordinate system that was introduced in \cite{mich} to make 
certain Killing directions in the transverse space manifest.  The
reverse procedure of T-dualisation and lifting was performed in
\cite{mich}, to give the pp-wave in $D=11$.

\subsection{AdS Penrose limits}

    Various AdS spacetimes can arise in M-theory as near-horizon
limits of M-brane intersections.  The M2-brane and M5-brane themselves
have near-horizon limits AdS$_4\times S^7$ and AdS$_7\times S^4$
respectively; these both have the same Penrose limit.  The resulting
pp-wave is maximally supersymmetric, and is the one obtained in
\cite{kg}, with
\bea
H&=&c_0-\ft19\mu^2\, (z_1^2 + z_2^2 + z_3^2) -\ft{1}{36} \mu^2\, (z_4^2 +
\cdots z_9^2)\,,\nn\\
\Phi_\3 &=& \mu\, dz^{123}\,.
\eea
(The constant $c_0$ was not included in the Penrose limit taken in
\cite{kg}, but it can easily be included, as was shown in
\cite{clppenrose} for any Penrose limit.)  There is no isometry along
any of the $z^i$ directions.  All the Killing spinors depend on $x^+$,
and so a dimensional reduction on the $x^+$ coordinate will result in
a type IIA supergravity solution that breaks all the supersymmetry.

     The Penrose limits of intersecting M-branes that give rise to AdS
structure were discussed in \cite{bur1.5}.  An M2/M5 brane
intersection gives rise to AdS$_3\times S^3\times T^4$ in the
near-horizon limit.  The Penrose limit is then given by
\bea
H&=& c_0 -\ft14 \mu^2 (z_1^2 + z_2^2 + z_3^2 + z_4^2)\,,\nn\\
\Phi_\3 &=& \mu\, (dz^{129} + dz^{349})\,.
\eea
There are in total 24 Killing spinors, of which 16 are the standard
Killing spinors, together with 8 supernumerary Killing spinors.  Out
of the 16 standard Killing spinors, 8 are independent of $x^+$.  All 8
supernumerary Killing spinors are independent of $x^+$, giving a total
of 16 Killing spinors independent of $x^+$. Thus the D0-brane brane
after reduction to type IIA will have 16 Killing spinors, but reduced
to 8 if the D0-brane charge $Q$ is turned on (since then the
supernumerary Killing spinors will already be lost in $D=11$).  The
solution has also isometries along the five $z^i$ coordinates with
$i=5,6,7,8,9$.  All 24 Killing spinors are independent of these
coordinate, and so they will all survive if the M-theory solution is
instead reduced on any of these five $z^i$ coordinates, giving a
pp-wave in type IIA.  In particular, if we reduce the solution on the
$z^9$ coordinate the pp-wave will have a constant (NS-NS) 3-form
$F_\3$, and in fact this pp-wave is itself the Penrose limit of the
NS1/NS5 system in type IIA.  If instead we reduce the $D=11$ solution
on $z_5$, $z_6$, $z_7$ or $z_8$, the ten-dimensional pp-wave is
supported by a constant R-R 4-form $F_\4$.

        M2/M2/M2 and M5/M5/M5 brane intersections give rise to
AdS$_2\times S^3$ and AdS$_3\times S^2$ respectively. Both have the
same Penrose limit, which can be written as
\bea
H&=& c_0-\ft14\mu^2\, (z_7^2 + z_8^2) - \mu^2\, z_9^2\,,\nn\\
\Phi_\3&=&\mu (dz^{129} + dz^{349} + dz^{569})\,,
\eea
In this case there are the 16 standard Killing spinors plus 4
supernumerary Killing spinors, giving a total of 20 in all.  All the
Killing spinors depend on $x^+$, and hence after reduction on $x^+$ to
type IIA the resulting D0-brane will have no supersymmetry.  There are
also isometries in the $z^i$ coordinates with $i=1,\ldots 6$.  Since
none of the Killing spinors depends on any of these coordinates, the
type IIA pp-wave that results from reducing instead on one of these
will have all 20 Killing spinors.

     The M2/M2/M5/M5 brane intersection system gives an AdS$_2\times
S^2$ in its near-horizon limit.  The Penrose limit is given by
\bea
H&=&c_0 - \mu^2\, (z_1^2 + z_6^2)\,,\nn\\
\Phi_\3 &=& \mu\, (dz^{123} + dz^{145} + dz^{246} + dz^{356})\,.
\eea
In this case, 4 out of the 16 standard Killing spinors are independent of
$x^+$.  Additionally, there are 4 supernumerary Killing spinors, which are
all $x^+$-independent.

\section{Matrix model and type II string actions}

\subsection{Type IIA string action}

    Whenever there is an isometry in any of the $z^i$ directions, the
$D=11$ pp-wave can be dimensionally reduced on such a coordinate, to
give a pp-wave in the type IIA theory.  In general, the resulting
pp-wave can have non-vanishing constant backgrounds both for the NS-NS
3-form and R-R 4-form. The precise details depend in the usual way on
the structure of $\Phi_\3$ in $D=11$.

   The Green-Schwarz action for the type IIA string in an arbitrary
bosonic background was derived, in component form up to and including 
second-order in the fermionic coordinates $\theta$, in
\cite{clps} (the form of the R-R couplings had previously 
been obtained schematically in \cite{tesyt}):
\bea
{\cal L}_2 &=& -\ft12 \sqrt{-h}\, h^{ij}\, \del_i X^\mu\, \del_j\,
X^\nu\, g_{\mu\nu} + \ft12\ep^{ij}\, \del_i\, X^\mu\, \del_j\, X^\nu\,
A_{\mu\nu} \nn\\
&&- \im\, \btheta\, \beta^{ij}\,
\Gamma_\mu\, D_j\theta\, \del_i X^\mu +\ft{\im}8 
\del_i X^\mu\, \del_j X^\nu\, \btheta\,\beta^{ij}\,\Gamma_{11}\,  
\Gamma_\mu{}^{\rho\sigma}\, \theta\,
F_{\nu\rho \sigma}  \label{type2alag}\\
&&-\ft{\im}{16} \del_i X^\mu \del_j X^\nu e^{\phi}\, \btheta 
\,\beta^{ij}\, \Big(\Gamma_{11}\,
\Gamma_\mu\, \Gamma^{\rho\sigma}\, \Gamma_\nu\, F_{\rho\sigma} + \ft1{12}
\Gamma_\mu\, \Gamma^{\rho\sigma\lambda\tau}\, \Gamma_\nu\,
F_{\rho\sigma\lambda\tau} \Big)\, \theta\,,\nn
\eea
where
\be
\beta^{ij} \equiv \sqrt{-h}\, h^{ij} -\ep^{ij}\, \Gamma_{11}\,,\qquad
D_i\theta \equiv  \del_i\theta + \ft14 \del_i X^\mu\,
\omega_\mu{}^{mn}\, \Gamma_{mn}\, \theta\,.\label{sdef}   
\ee
The field strengths are given by
\be
F_\4 = dA_\3 - A_\1\wedge dA_\2\,,\qquad F_\3= dA_\2\,,\qquad 
F_\2 = dA_\1\,.
\ee

   The fermionic coordinates $\theta$ are non-chiral, and can be
written as $\theta=\pmatrix{\theta^1\cr \theta^2}$, where
$\Gamma_{11}\, \theta =\pmatrix{\theta^1\cr -\theta^2}$.  In a
notation adapted to the passage to light-cone gauge, we can introduce
world-sheet Dirac matrices $\varrho_i$, with $\varrho_0=-\im\,
\tau_2$, $\varrho_1=\tau_1$ and $\varrho_2=\tau_3$, where $\tau_i$ are
the Pauli matrices.  The $\varrho_i$ act on the upper and lower 16
components $\theta^1$ and $\theta^2$ of the column vector
$\Psi=\pmatrix{\theta^1\cr\theta^2}$, and $\varrho_2$ is the chirality
operator.  The conjugate
spinor in this notation is then $\bar\Psi=\Psi^\dagger\, \Gamma_0\, 
\varrho_0$,
and we therefore have the ``dictionary'' $\btheta\, {\cal O}\,
\theta \longrightarrow -\bar\Psi\, \varrho_0\, {\cal O}\, \Psi$ and
$\btheta\, \Gamma_{11}\, {\cal O}\, \theta \longrightarrow \bar\Psi\,
\varrho_1\, {\cal O}\, \Psi$, where ${\cal O}$ is any matrix or operator
constructed from the $\Gamma_i$ matrices.
In the light-cone gauge, where $X^+=\tau$, $\Gm\, \theta=0$ and
$\sqrt{-h}\, h^{ij}=\eta^{ij}$, the fermionic part of the type IIA
Green-Schwarz action (\ref{type2alag}) therefore becomes
\be
{\cal L}_F = \im\, \bar\Psi\, \Gp\, \not\!\! D\, \Psi 
- \ft{\im}4 \bar\Psi\, \varrho_1\,\Gp\, \not\!\! F_\3\, \Psi
- \ft{\im}4 e^{\phi}\, \bar\Psi\, \Gp\, (\varrho_1\, \not\!\! F_\2
- \varrho_0\, \not\!\! F_\4)\, \Psi\,,\label{2arr}
\ee
where we have defined
\be
\not\!\! F_\2\equiv \Gamma^i\, F_{+i}\,,\qquad
\not\!\! F_\3\equiv \ft12 \Gamma^{ij}\, F_{+ij}\,,\qquad
\not\!\! F_\4\equiv \ft16 \Gamma^{ijk} \, F_{+ijk}\,.
\ee
In the pp-wave backgrounds we are considering here, the world-sheet
Dirac operator $\not\!\! D$ just reduces to $\not\!\del$ in the light-cone
gauge.

   If the dimensional reduction from the $D=11$ pp-wave to $D=10$ is
performed on a Killing direction $z^i$ whose differential $dz^i$ does
not appear in the expression for $\Phi_\3$ in $D=11$, then the
solution in $D=10$ is a pp-wave with only the R-R 4-form $F_\4$ as a
source.  This situation can be achieved for any $\Phi_\3$ contained
within Case 2, provided that one reduces on the $z_8$ or $z_9$
coordinate.  Since Case 1 is encompassed by Case 2 (after appropriate
coordinate relabellings) in all situations except where all four
$m_\a$ are non-vanishing, it is only in this last circumstance that
one is forced into a dimensional reduction in which the differential
$dz^i$ of the reduction coordinate is present in $\Phi_\3$ in $D=11$.
Of course in other cases too, one may choose to perform the
dimensional reduction on such a coordinate $z^i$, provided that the
associated coefficient $\mu_i$ in the quadratic metric function $H$
vanishes, implying that $\del/\del z^i$ is a Killing vector.

  Let us first consider the case where the differential $dz^i$ of the
reduction coordinate $z^i$ does not appear in $\Phi_\3$ in $D=11$.  It
then follows from (\ref{type2alag}) and (\ref{2arr}) that after
choosing the light-cone gauge, the associated type IIA string action
will be\footnote{See also \cite{hysh} for a related discussion of 
the type IIA Green-Schwarz action in a gravitational wave background.}
\bea
{\cal L} =
\sum_{i=1}^8 (\ft12 \dot z_i^2 -\ft12 z_i'^2 -\ft1{2}\mu_i^2
z_i^2) + \bar\Psi (\im \not\! \del + 
\ft{1}4\, \mu\,\varrho_0\,  W)\, \Gp\, 
\Psi\,.
\eea
Thus the masses of the fermions will be given by the eigenvalues of $W$.

   Suppose now that we instead perform a reduction on a coordinate $z^i$
whose differential $dz^i$ does appear in $\Phi_\4$ in $D=11$.  We now find
that $F_\4$ in $D=11$ reduces as
\be
F_\4 \longrightarrow F_4 + dz\wedge F_\3
\ee
in $D=10$, where $z$ is the reduction coordinate.  We now have the 
non-vanishing NS-NS field $F_\3$ in the type IIA background, together,
possibly, with a non-vanishing R-R 4-form $F_\4$.   If the reduction of
the 3-form $\Phi_\3$ is written as $\Phi_\3\longrightarrow \Phi_3
+ dz\wedge \Phi_2$, and if we define  
\be
Y \equiv \ft{\im}2 \Phi_{ij}\, \Gamma^{ij}\,,
\ee
then it follows from (\ref{type2alag}) and (\ref{2arr}) that after
choosing the light-cone gauge, we shall obtain the string action
\bea
{\cal L} =
\sum_{i=1}^8 (\ft12 \dot z_i^2 -\ft12 z_i'^2 +\mu\,  z_i'\, B_i
-\ft1{2}\mu_i^2
z_i^2) 
+ \bar \Psi (\im \not \!\del + \ft{1}4\, \mu\, \varrho_0\, W + 
   \ft{1}4 \, \mu \, \varrho_0\, Y)\, \Gp\, 
\Psi\,.
\eea
Here, $B_i$ denotes the components of the 1-form $B_\1$ whose exterior
derivative gives $\Phi_\2=dB_\1$.  Since $\Phi_\2=\ft12 \Phi_{ij}\,
dz^i\wedge dz^j$ where $\Phi_{ij}$ are constants, we may therefore
take $B_i$ to be given by
\be
B_i = \ft12 \Phi_{ji}\, z^j\,.
\ee
Thus the string action in this case is given by
\bea
{\cal L} &=&
\sum_{i=1}^8 (\ft12 \dot z_i^2 -\ft12 (z_i'-\ft12 \mu\, \Phi_{ij}\, z_j)^2 
-\ft1{2}\mu_i^2 z_i^2) +\ft18 \mu^2 \Phi_{ik}\, \Phi_{jk}\, z^j\,z^k  
\nn\\
&&+ \bar\Psi (\im \not\! \del +\ft{1}4\, \mu\, \varrho_0\, W 
+ \ft{1}4 \, \mu\, \varrho_0\,  Y)\, \Gp\, 
\Psi\,.
\eea
Thus the boson masses, as well as the fermion masses, are modified by the
presence of the NS-NS 3-form field.  This is a generalisation of a result
obtained in \cite{bermalnas}.

  As an example, let us consider the pp-wave in $D=11$ 
resulting from taking $\Phi_\3$ to be given by Case 1, as in (\ref{case1}).
After dimensional reduction on the coordinate $z_9$, which will be a
Killing direction provided that 
\be
m_1+m_2+m_3+m_4=0
\ee
(see (\ref{case1mu9})), the type IIA light-cone action will be given by
\bea
{\cal L} &=&
\sum_{i=1}^8 (\ft12 \dot z_i^2 -\ft12 (z_i'-\ft12 \mu\, \Phi_{ij}\, z_j)^2 
-\ft1{2}\mu_i^2 z_i^2) +\ft18 \mu^2 [m_1^2\, (z_1^2+z_2^2) + \cdots 
+ m_4^2\, (z_7^2 + z_8^2)] \nn\\
&&  
+  \bar\Psi (\im \not\! \del + \ft{1}4 \, \mu\, \varrho_0\,  Y)\, \Gp\, 
\Psi\,,
\eea
with $Y$ given by
\be
Y= \im\, m_1\, \Gamma_{12} + \im\, m_2 \, \Gamma_{34} + 
\im\, m_3\, \Gamma_{56} + \im\, m_4\, \Gamma_{78}\,.
\ee
(The matrix $W$ is absent here, since all terms in $\Phi_\3$ in
$D=11$ involved a factor $dz_9$.  Thus the $D=10$ background is 
purely NS-NS in this example.)

   Note that the choice of gauge for writing $B_\1\equiv B_i\, dz^i$
is not unique.  In this example we could, for instance, choose,
instead of writing it in the ``symmetrical'' gauge
\be
B_\1 = \ft12 \Phi_{ij}\, z^i\, dz^j = \ft12m_1\, (z_1\, dz_2-z_2\, dz_1) 
  + \cdots + \ft12 m_4\, (z_7\, dz_8-z_8\, dz_7)\,,
\ee
to write it in the ``asymmetrical'' form
\be
B_\1 = m_1\, z_1\, dz_2 + m_2\, z_3\, dz_4 +m_3\, z_5\, dz_6 
+ m_4\, z_7\, dz_8
\ee
In this choice of gauge we would instead obtain the string action
\bea
{\cal L} &=&
\sum_{i=1}^8 (\ft12 \dot z_i^2 -\ft12 (z_i'-\ft12 \mu\, \Phi_{ij}\, z_j)^2 
-\ft1{2}\mu_i^2 z_i^2) +\ft12 \mu^2 [m_1^2\, z_1^2 + m_2^2 z_3^2 +  
m_3^2\, z_5^2 + m_4^2\, z_7^2] \nn\\
&&  
+ \bar\Psi (\im \not \!\del + \ft{1}4 \, \mu \, 
\varrho_0\, Y)\, \Gp\,  \Psi\,.
\eea
Of course the different gauge choices just change the action by a
total derivative, and so they are equivalent in the closed string sector.

\subsection{Type IIB string action}

   Many of our examples can be T-dualised to pp-waves in type IIB
theory when there are two $\mu_i$ that vanish.  In some cases, when the
type IIA pp-wave is supported only by the $F_\3$, the solution is also
valid in type IIB theory supported by the NS-NS $F_\3$ or the R-R
$F_\3$, or both using S-duality rotation.  Thus in this section, we
consider the light-cone type IIB string action in such a background.

   In \cite{clps}, the Green-Schwarz action for the type IIB string
in an arbitrary bosonic background was derived, giving
all terms up to and including quadratic order in the fermionic
coordinates.  In the notation of \cite{clps}, the two Majorana-Weyl
fermions were denoted by $\theta^1$ and $\theta^2$.  If we put these in
a column vector $\theta\equiv \pmatrix{\theta^1\cr\theta^2}$, and define
world-sheet Dirac matrices $\varrho_i$ by $\varrho_0= -\im\, \tau_2$,
$\varrho_1= \tau_1$ and $\varrho_2=\tau_3$, where $\tau_i$ are the Pauli
matrices, then we find the following ``dictionary'' for converting
the notation in \cite{clps} to the one we wish to use here.  For any
matrix or operator ${\cal O}$ constructed from the target-space
Dirac matrices, we shall have
\bea
&&\btheta\, \varrho_0\, {\cal O}\, \theta=
 -\bar\theta^1\, {\cal O}\,\theta^1 - \btheta^2\, {\cal O}\, \theta^2\,,
\qquad
\btheta\, \varrho_1\, {\cal O}\, \theta=
 \bar\theta^1\, {\cal O}\,\theta^1 - \btheta^2\, {\cal O}\, \theta^2\,,
\nn\\
&&
\btheta\, {\cal O}\, \theta= 2 \bar\theta^{[1}\, {\cal O}\, \theta^{2]}
\,,\qquad
\btheta\, \varrho_2\, {\cal O}\, \theta=
    -2   \bar\theta^{(1}\, {\cal O}\, \theta^{2)} \,,
\eea
where the conjugate of $\theta$ is defined by
$\btheta=\theta^\dagger\, \Gamma_0\, 
\varrho_0= (-\bar\theta^2,\bar\theta^1)$.
Substituting into equation (3.29) of \cite{clps} (in the updated
v2. where a minor typographical error has been corrected and conventions
adjusted), the type IIB Green-Schwarz action up to $O(\theta^2)$ is
given by
\crampest
\bea
{\cal L} &=& -\ft12 \sqrt{-h}\, h^{ij}\, \del_i X^\mu\, \del_j X^\nu\,
           g_{\mu\nu} + \ft12\ep^{ij}\, \del_i X^\mu\, \del_j X^\nu\,
        B_{\mu\nu} \nn\\
&&+ \im\, \del_i X^\mu\, \btheta\, \gamma^{ij}\, \varrho_0\, 
\Gamma_\mu\, D_j\,\theta
-\ft{\im}8 \del_i X^\mu\, \del_j X^\nu\, \btheta\, \gamma^{ij}\, 
\varrho_1\, \Gamma_\mu{}^{\rho\sigma}\,
\theta\, G_{\nu\rho\sigma} \label{2baction}\\
&& + \ft{\im}{8} e^{\phi}\, \del_i X^\mu\, \del_j X^\nu \,
\btheta\, \gamma^{ij}\, \Gamma_\mu\,
[ \Gamma^\rho\, \del_\rho\chi + \ft16\varrho_2\,  
\Gamma^{\rho_1\rho_2\rho_3}\, F_{\rho_1\rho_2\rho_3}
+\ft1{240} \Gamma^{\rho_1\cdots \rho_5}\,  
F_{\rho_1\cdots \rho_5}] \,\Gamma_\nu\,  \theta \,,\nn
\eea 
\uncramp
where $G_\3=dB_\2$ is the NS-NS 3-form, $\phi$ and $\chi$ are the
dilaton and axion, and $F_\3$ and $F_\5$ are the R-R 3-form and
self-dual 5-form, and we have defined
\be
\gamma^{ij}\equiv \sqrt{-h}\, h^{ij} - \ep^{ij}\, \varrho_2\,.
\ee

   In the light-cone gauge, $X^+=\tau$, $\theta=\Psi$ with $\Gm\, \Psi=0$, and
$\sqrt{-h}\, h^{ij} =\eta^{ij}$, we therefore find that the fermionic
part of the action becomes
\be
{\cal L}_F = \im\, \bar\Psi\, \Gp \not \!\! D\, \Psi + \ft{\im}4
\bar\Psi\, \varrho_1\,\Gp \not \!\! G_3\, \Psi
-\ft{\im}{4}\, e^\phi\, \bar\Psi \, \Gp\, (\del_+\chi
\,\, +\varrho_2\, \not \!\! F_\3 + \ft12 \not \!\! F_\5)\, \Psi\,,
\label{2blightcone}
\ee
where we have defined
\be
\not \!\! G_\3 \equiv \ft12 \Gamma^{ij}\, G_{+ij}\,,\qquad
\not \!\! F_\3 \equiv \ft12 \Gamma^{ij}\, F_{+ij}\,,\qquad
\not \!\! F_\5 \equiv \ft1{24} \Gamma^{ijk\ell}\,
   F_{+ijk\ell}\,.\label{fslash}
\ee
Note that for all the pp-waves, it follows from (\ref{spincon}) that
$\not \!\! D$ is just $\not\! \del$.  

      To apply this action to our examples, let us first consider Case
2, with $\mu_8=\mu_9=0$.  This example can be T-dualised to type IIB,
where it gives rise to pp-waves of the sort described in
\cite{clppenrose}.  In these cases the pp-wave is supported
only by the self-dual R-R 5-form.  The action was obtained in 
\cite{clppenrose}.

    For another example consider Case 1, with all four of the $m_\a$
non-vanishing, but chosen so that $\mu_9=0$, thus permitting a reduction
in the $z_9$ direction to give a type IIA solution.  The solution is then
supported purely by the NS-NS 3-form.  There is in general no further
isometry direction among the remaining $z_i$ coordinates that could
allow us to perform a T-duality transformation.  However, since only
the NS-NS 3-form field is involved, we can clearly take the identical
configuration and view it as a solution instead of the type IIB
theory, again supported by the NS-NS 3-form.  Having done so, we can
then choose to perform an S-duality transformation, thereby
introducing a non-vanishing R-R 3-form as well (or instead).  It is then
straightforward to see from (\ref{2blightcone}) and (\ref{fslash}), 
together with imposing the light-cone gauge in the bosonic part of 
(\ref{2baction}), that these type IIB pp-wave backgrounds will have
a light-cone string action given by
\be
{\cal L}=
\sum_{i=1}^8 (\ft12 \dot z_i^2 -\ft12 (z_i'-\ft12 B_{ij}\, z^j)^2 
-\ft1{2}\mu_i^2
z_i^2) + \bar\Psi (\im\not\! \del + 
  \ft{\im}4 \varrho_1\, \not\!\! G_\3 
 - \ft{\im}4 \varrho_2\, \not\!\! F_\3)\, \Gp\, \Psi\,,
\ee
where $\not\!\! G_\3$ and $\not\!\! F_\3$ are the NS-NS and R-R 3-form
contributions respectively.

\subsection{Matrix model action}

     There are many examples in our general discussion where all the
coefficients $\mu_i$ in the metric function $H$ are non-vanishing,
implying that the pp-wave is intrinsically eleven-dimensional.  In
these cases, the system is best described by a D0-brane
action. Namely, one can perform a DLCQ compactification
\cite{matrix1}-\cite{matrix4} along the light-cone coordinate $x^-
\equiv x^-+ 2\pi R$, and consider the sector with momentum
$2p^+=-p^-=N/R$.  The dynamics of this sector is then described by a
$U(N)$ matrix model with the strength of interactions governed by
$g\sim 2R$.  The procedure as it applies to the case of the Penrose
limit of AdS$_4\times S^7$ or AdS$_7\times S^4$ was given in
\cite{bermalnas}.  The form of the action for the general, constant
$W$, as studied in this paper, can be derived along the same lines,
and is structurally of the same form.  This is due to the fact that
the 4-form field strength enters the D0-brane particle action in the
light-cone gauge (\ie the $U(N)$ matrix model) only through
$W=\ft{\im}{6} \Phi_{ijk}\,\Gamma^{ijk}$.  The form of the action is thus
given by
\bea
L&=&\sum_{i=1}^9 [(\dot X^i)^2 -\mu_i^2 (X^i)^2] + \Psi^T\, \dot \Psi +
\ft{\im}4\, \mu\, \Psi^T\, W\, \Psi\nn\\
&&- \ft23\mu\,  g\, \sum_{i,j,k=1}^9 {\rm Tr}(X^i\, X^j\, X^k)\,\Phi_{ijk} +
2g^2\, {\rm Tr}([X^i, X^j]^2) + 2{\rm i}g\, 
{\rm Tr}(\Psi^T\,\Gamma^i\,[\Psi, X^i])\,.
\eea
Note that in addition to the standard matrix-model interactions there
are also the fermionic and bosonic mass terms, and additionally the
term tri-linear in $X^i$ that is related to the Myers effect \cite{myers}.

   The supersymmetry of this quantum mechanical matrix model fixes the
coefficients in front of the fermionic mass terms and the interaction
terms in the same way as it was derived for the special case of
$W=\Phi_{123}\,\Gamma^{123}$ in \cite{bermalnas}. Indeed, the existence
of supersymmetry is dictated by the existence of the supernumerary
Killing spinors.  In fact, the supersymmetry transformation 
parameter is exactly the supernumerary Killing spinor:
\bea \delta X^i &=& \Psi\, \Gamma^i\, \epsilon\,,\nn\\ 
\delta \Psi &=& \Big(\dot X^i\, \Gamma_i + \mu\, X^i\, 
(-\ft14 W\, \Gamma_i + \ft1{12}
\Gamma_i\, W) + {\rm i}\, g\, [X^i, X^j]\, \Gamma_{ij} \Big)\,
\epsilon\,,\nn\\ 
\epsilon &=& e^{\mu\, W\, t}\, \epsilon_0\,.  
\eea 
The case where $W=\Gamma_{123}$ was given in \cite{bermalnas}. In that
case, the system is fully supersymmetric, and hence $\epsilon_0$ is an
arbitrary constant spinor.  Furthermore, since $W$ has no zero
eigenvalues in that example, all the supersymmetry parameters
$\epsilon$ are time-dependent.  

   For the more general $W$'s that we have considered in this paper,
$\epsilon_0$ is subject to further projection constraints, in
accordance with the supernumerary Killing spinors.  In our more
general cases $W$ can annihilate $\epsilon_0$, implying that
$\epsilon$ is then time-independent. In such an example, the pp-wave
can also be reduced to give rise to a pp-wave in type IIA, thus giving
an exactly-solvable string action.  The existence of two routes, one
corresponding to the matrix model of the D0-particle action, and the
other corresponding to the free massive Type IIA string action,
therefore suggests that these are dual descriptions of the theory when
the background is of this particular type.

\section{Conclusions}

    In this paper, we studied a general class of supersymmetric
pp-waves in M-theory, by turning on constant 3-forms, motivated by
the orbifold (flat) limit of a special holonomy transverse space for
the pp-wave.  These 3-forms fall into two classes, one motivated
by the K\" ahler form of the eight-dimensional special holonomy
transverse space, and the other motivated by the associative 3-form of
a seven-dimensional transverse space of $G_2$ holonomy.

 This general class of pp-waves encompass the Penrose limits of
AdS$_p\times S^q$ with $(p,q)=(4,7),\, (7,4),\, (3,3),\, (3,2),\,
(2,3),\, (2,2)$ which are associated with the near horizon limits of
the M2-brane, M5-brane, and M2/M5, M5/M5/M5, M2/M2/M2 and M2/M2/M5/M5
intersections, respectively. In addition this general class contains
many additional examples of pp-waves that are do not correspond to any
known Penrose limit.

   We focused on the study of the target space supersymmetry. In
addition to 16 ``standard'' Killing spinors that always arise, we
determined the conditions under which additional ``supernumerary''
Killing spinors appear. We also analysed the conditions under which
the Killing spinors are independent of the light cone $x^+$
coordinate, or of one or more of the nine transverse
coordinates. These conditions determine whether the reduction of the
M-theory pp-waves to type IIA supergravity, and subsequent
T-dualisation to type IIB, remain supersymmetric.

    Since $x^+$ is always a Killing direction the M-theory pp-wave can
always be reduced on this coordinate, leading to a D0-brane
configuration of the type IIA theory. Its world-particle action
corresponds to a DLCQ description of a matrix-theory action for
M-theory, with unbroken supersymmetry governed by $x^+$-independent
supernumerary supersymmetries of the M-theory pp-wave background.

  On the other hand the independence of a Killing spinor on a
transverse coordinate allows for a reduction on this coordinate down
to a supersymmetric type IIA pp-wave. The light cone string actions in
these backgrounds correspond to exactly-solvable free massive string
theories, and again the supernumerary supersymmetries play a key role
in determining the supersymmetry of the string action.

\section*{Note Added}

   We have updated the discussion of the type IIA and type IIB string
actions in this version of the paper, taking into account the
corrections and improvements in v2.~of \cite{clps}.  There was only
one minor typographical in the type IIB action in the earlier version
of \cite{clps}, but there were various inelegant notations and
conventions, all of which have now been changed, and these changes are
incorporated in this version of the present paper.  A detailed
discussion of the changes is given in the Addendum section in v2.~of
\cite{clps}. We are grateful to Kelly Stelle for discussions leading
to these improvements.

\section*{Acknowledgements}

   We are grateful to Gary Gibbons, Jim Liu and Justin
V\'azquez-Poritz for conversations.  H.L.~and C.N.P.~are grateful to
UPenn for hospitality and financial support during the course of this
work.

\end{document}